\documentclass[10pt,aps,prb,twocolumn,superscriptaddress,floatfix,english,twocolumn,amsmath,amssymmb,longbibliography]{revtex4-2}
\usepackage{amsfonts}
\usepackage{babel}
\usepackage{bm}
\usepackage{color}
\usepackage{comment}
\usepackage{dcolumn}
\usepackage{enumerate}
\usepackage{epsf}
\usepackage{subfigure}
\usepackage{epsfig,psfrag}
\usepackage{esint}
\usepackage{float}
\usepackage[T1]{fontenc}
\usepackage{graphicx}
\usepackage[normalem]{ulem}
\usepackage{latexsym}
\usepackage{physics}
\usepackage{relsize}
\usepackage[final]{hyperref} 
\hypersetup{
	colorlinks=true,       
	linkcolor=blue,        
	citecolor=blue,        
	filecolor=magenta,     
	urlcolor=blue         
}
\usepackage{soul}
\usepackage{xcolor}

\newcommand{\tx}{x}
\newcommand{\tL}{L}
\newcommand{\tB}{B}

\newcommand{\co}{c}

\begin{document}
	
\title{Channel-selective magnetic filtering in a nodal-line semimetal}

\author{Hironmoy Pratihar}
\email{hironmoy20@iitk.ac.in}
\affiliation{Department of Physics, Indian Institute of Technology - Kanpur, Kanpur 208016, India}

\author{Alessandro De Martino}
\email{ademarti@city.ac.uk}
\affiliation{Department of Mathematics, City St George's, University of London, London EC1V 0HB, United Kingdom}

\author{Arijit Kundu}
\email{kundua@iitk.ac.in}
\affiliation{Department of Physics, Indian Institute of Technology - Kanpur, Kanpur 208016, India}

\date{\today}
	
\begin{abstract}
\noindent We study quantum transport through a magnetic barrier in a nodal-line semimetal. 
When the Fermi energy lies near the nodal ring, the Fermi surface has toroidal geometry.
Each cross-section in a plane parallel to the nodal ring consists of two concentric contours, 
inner and outer, carrying distinct transport channels. 
We show that a magnetic barrier resolves these two channels: 
because the contours enclose different momentum-space areas, 
they accommodate the field-induced transverse-momentum shift unequally, 
and the inner channel is cut off at a weaker barrier strength than the outer.
Using a two-band effective Hamiltonian and a wave-function matching approach, we obtain closed-form, 
channel-resolved transmission amplitudes. Over a finite window of barrier strength the inner contour 
is fully blocked while the outer still transmits, so the barrier acts as a channel-selective filter. 
This sequential quenching shapes the two-terminal conductance, which decreases with barrier strength 
as the two channels close in turn and terminates once the outer channel is cut off, 
providing experimentally accessible fingerprints of the toroidal Fermi surface of a nodal-line semimetal.
\end{abstract}

\maketitle

\section{Introduction}
\label{Sec:introduction}

Three-dimensional nodal-line semimetals (NLSs) 
are topological semimetals in which the conduction and valence bands cross along
one-dimensional manifolds in the Brillouin
zone ~\cite{burkov2011topological,fang2016topological,bian2016topological,chan20163,yang2018,yang2022quantum},
in contrast to Weyl or Dirac semimetals, where band crossings occur at
isolated points~\cite{armitage2018weyl}. These extended
degeneracies are protected by crystalline symmetries such as
mirror reflection or nonsymmorphic operations~\cite{fang2016topological,schoop2016nodal}. 
Candidate materials include $\mathrm{Ca}_3\mathrm{P}_2$~\cite{xie2015new,chan20163},
ZrSiS~\cite{schoop2016nodal},
AlB$_2$~\cite{takane2018observation},
and PbTaSe$_2$~\cite{bian2016topological}.
Among these, $\mathrm{Ca}_3\mathrm{P}_2$ provides a useful reference system: its nodal ring lies 
close to the Fermi level and the small atomic numbers render 
spin–orbit coupling negligible~\cite{chan20163}. We therefore use parameters appropriate
to  $\mathrm{Ca}_3\mathrm{P}_2$ for numerical estimates throughout.

In a NLS, when the Fermi energy lies close to the nodal ring, 
the constant-energy surface has the geometry of a torus.
A cross-section in a plane parallel to that of the nodal ring cuts it in two 
concentric contours, so that at fixed energy and conserved transverse momenta, 
there are two distinct longitudinal wavevectors, corresponding to an outer and an inner contour~\cite{yang2022quantum}.
For each propagation direction there is one state on each contour, 
so a fixed-energy slice supports two co-propagating channels, 
a feature with no counterpart in Weyl or Dirac systems, 
where each node contributes a single Fermi-surface contour.
This two-contour structure is a direct consequence of the toroidal geometry of
the Fermi surface. It manifests in unusual Landau-level spectra~~\cite{rhim2015landau,molina2018surface} 
and magneto-optical responses~\cite{yang2022quantum}, and has recently been
probed in transport experiments on ZrTe$_5$, where fields parallel to
the nodal-line plane induce flat bands and nonlinear
magnetotransport~\cite{Wang2023}.
The related problem of interface and junction physics of NLSs has recently 
been explored in some detail: drumhead-derived states at NLS boundaries and their
smooth-interface dispersion have been characterized~\cite{Buccheri2024,
Rudi2024}, and ballistic transport across NLS interfaces exhibits
resonant angles of perfect transmission and radially inhomogeneous
current profiles~\cite{Rudi2024}. Here we show that a magnetic barrier
probes a complementary aspect of the same toroidal Fermi surface, its
two-contour cross-section, through the transport channels it supports.

Magnetic barriers are a natural way to probe such momentum-space structure. 
In two-dimensional electron gases they produce wavevector-dependent tunneling 
and Landau-level bending near barrier edges~\cite{matulis1994wave}. 
In graphene, they lead to wavevector filtering~\cite{demartino2007magnetic,ramezani2008direction}, 
snake states~\cite{ghosh2008electron,oroszlany2008theory}, 
and magnetic waveguiding~\cite{demartino2007magnetic}. 
In three-dimensional Weyl semimetals related magnetic-barrier geometries
generate perfect-transmission rings and selective momentum
filtering~\cite{yesilyurt2016klein}.
The common mechanism is that a magnetic barrier shifts the kinetic transverse momentum 
of transmitted carriers proportional to the product of the field strength and
the barrier width, rather than shifting all carrier energies as an
electrostatic barrier does.

To our knowledge, previous studies of barrier tunneling in nodal-line 
semimetals have focused on electrostatic barriers~\cite{Khokhlov2018,Guan2018}, 
where Klein-like phenomena and magic-angle resonances arise. 
A magnetic barrier acts instead on the conserved transverse momentum: 
taken on its own, each Fermi contour behaves much like the single Dirac cone of
graphene, with the barrier shifting its transverse momentum and
transmitting only states that remain within the
contour~~\cite{ramezani2008direction,demartino2007magnetic}. The toroidal Fermi surface,
however, offers two such contours, and the barrier is sensitive to
this structure. Because the inner and outer contours enclose
different momentum-space areas, they accommodate the transverse shift
unequally: the smaller inner contour is cut off at a weaker barrier
strength than the outer. There is consequently a window of barrier
strength in which the inner channel is completely blocked while the
outer still transmits, so that the magnetic barrier acts as a
channel-selective filter that exposes the two-contour
structure of the toroidal Fermi surface. 

In this paper we develop a complete analytical treatment of this
problem, using a two-band effective Hamiltonian appropriate for
materials such as Ca$_3$P$_2$. Expressing the magnetic-region wave
functions in terms of parabolic-cylinder functions and matching the wave
functions across the barrier yields closed-form transmission
amplitudes for each channel, from which we compute the two-terminal
conductance. Because the inner and outer contours quench sequentially,
the conductance versus field decreases as the inner channel closes 
and then terminates when the outer channel is cut off. 
This behavior, and the underlying channel-selective filtering, 
is an experimentally accessible fingerprint 
of the toroidal Fermi surface characteristic of a NLS.

The paper is organized as follows. In Sec.~\ref{Sec:Model} we introduce
the two-band Hamiltonian for the nodal-line semimetal and 
the magnetic-barrier geometry. In Sec.~\ref{Sec:Eigensystem} we derive the eigensystem in the
non-magnetic and magnetic regions. In Sec.~\ref{Sec:Transport} we establish
the critical condition for nonvanishing transmission, derive the
channel-resolved scattering amplitudes, and discuss the numerical results 
for transmission, angular dependence, and
conductance. We conclude with a summary and an outlook in Sec.~\ref{Sec:Conclusions}.

\section{Model}
\label{Sec:Model}

We consider a NLS in the presence of an inhomogeneous
magnetic field perpendicular to the plane of the nodal ring. 
We choose coordinates such that the nodal ring lies in the $xy$-plane while the
magnetic field points along $z$ and varies only along $x$. Near the nodal line, 
the NLS is described by the minimal two-band Hamiltonian~\cite{burkov2011topological, chan20163, yang2022quantum} 
(we set $\hbar =1$ throughout)
\begin{equation}
    \mathcal{H} = -i v\partial_z \tau_y
    + \lambda \left(\Pi_x^2 + \Pi_y^2 - k_0^2\right) \tau_z,
    \label{eq:H_dimful}
\end{equation}
where $\tau_{y,z}$ are Pauli matrices in orbital space, $v$ is the
velocity along $z$, $\lambda$ controls the in-plane band curvature, 
$k_0$ is the nodal-ring radius, 
and $\boldsymbol{\Pi} = -i\boldsymbol{\nabla} + e\mathbf{A}$ the
gauge-covariant momentum, with $-e$ the electron charge.
We neglect the small Zeeman coupling and a particle-hole-symmetry breaking 
term proportional to the identity, 
which is generically present unless a chiral symmetry forbids it. 
For numerical estimates, we use parameters appropriate to
$\mathrm{Ca}_3\mathrm{P}_2$, where the two relevant bands originate from P $p$-orbitals 
and Ca $d$-orbitals: $v\approx 2.50$\, eV\,\AA,
$\lambda \approx 4.34$~eV\,\AA$^2$ and $k_0\approx 0.206$\,\AA$^{-1}$~\cite{chan20163}.

For the field-free Hamiltonian, the constant-energy surface has toroidal topology when
$|E|<\lambda k_0^2$. Throughout this work, we focus on this regime, 
in which each fixed-$k_z$ cross-section generically contains inner and outer contours.

\begin{figure}[t]
    \centering
    \includegraphics[width=0.5\textwidth]{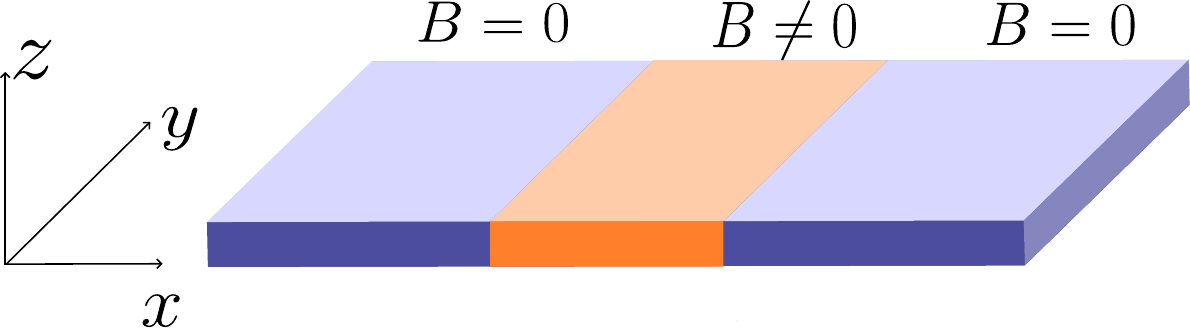}
    \caption{Schematic of the NLS magnetic-barrier geometry.
    The nodal ring lies in the $x$--$y$ plane and the magnetic field is nonzero
    only within the strip $0\le\tx\le\tL$. The system remains translationally 
    invariant along the transverse $y$ and $z$ directions.}
    \label{fig:setup}
\end{figure}

We model the magnetic barrier as a uniform magnetic field $B$ confined to a
region of width $L$ along $x$ (see Fig.~\ref{fig:setup}(a)). 
In the Landau gauge, the vector potential can be chosen as $\mathbf{A} = A_y(x) \, \hat{y}$, 
where
\begin{equation}
    A_y(x) =
    \begin{cases}
        0   & x < 0\\
        Bx  & 0 \le x \le L\\
        BL  & x > L
    \end{cases}.
\end{equation}
The step profile is an idealization of physical barrier edges of finite width. It should capture
the qualitative physics provided the edge-smoothing length is small compared with the barrier width 
and with the length scales over which the scattering states vary appreciably. 
Smooth edges may modify the transmission quantitatively, but not the channel-selective cutoff mechanism discussed
in this paper.

Translation invariance along $y$ and $z$ allows us to write the wave
function as $\Psi(x,y,z) = \psi(x) \, e^{i(k_y y + k_z z)}$, reducing
the problem to an effectively one-dimensional one for $\psi(x)$, with Hamiltonian
\begin{equation}
    \mathcal{H} = v k_z\tau_y
    + \lambda \left( - \partial_x^{2} + \left( k_y +  e A_y(x) \right)^2 - k_0^2\right) \tau_z.
\end{equation}
We introduce the magnetic length associated with a reference field $B_0$, 
$\ell_0= \sqrt{1/(eB_0)}$, and measure energies in units of $E_0 = \lambda/\ell_0^2$.
Dimensionless variables are defined by
\begin{equation}
\mathbf{q} = \ell_0 \mathbf{k}, \quad
\tilde{x} = \frac{x}{\ell_0}, \quad 
\tilde{L} = \frac{L}{\ell_0},\quad
\tilde{B} = \frac{B}{B_0}.
\end{equation}
In these units, the dimensionless Hamiltonian $ \tilde{\mathcal{H}}= \mathcal{H}/E_0$ 
reads
\begin{equation}
 \tilde{\mathcal{H}} = c q_z\tau_y
    + \left( -\partial_{\tilde{x}}^{2} + \left(q_y + \tilde B \tilde x\right)^2 - q_0^2\right)\tau_z.
    \label{eq:H_dimless}
\end{equation}
in the magnetic region, while the field-free form follows setting $\tilde B=0$. 
We have defined $c=v\ell_0/\lambda$ and $q_0=\ell_0k_0$. 
In the dimensionless variables introduced above, the toroidal regime corresponds to $|E|<q_0^2$.

Throughout this paper,  
we use the reference value $B_0=1$\,T, so that $\ell_0 \simeq 26$\,nm.
Then for the parameters of $\mathrm{Ca}_3\mathrm{P}_2$, we find $E_0 \approx 0.07$~meV, $c\approx 148$, $q_0 \approx 53$.
In the figures below, we shall use the $\mathrm{Ca}_3\mathrm{P}_2$ values of $q_0$ and $c$, 
while $B$, $L$, and $E$ are chosen for illustrative clarity. 
From here on we work exclusively in dimensionless units and drop the tildes, 
writing $B$, $L$, $x$ in place of $\tilde B$, $\tilde L$, $\tilde x$.

Magnetic barriers of the type considered here can be realized by
depositing a ferromagnetic stripe on the surface of the sample and
magnetizing it with an external field.
The perpendicular component of the fringe field generated by the stripe then acts as
a localized magnetic barrier on the underlying carriers, as demonstrated 
for two-dimensional electron gases \cite{nogaret2000electron,carmona1995two,Nogaret2010review}. 
Applying this to a nodal-line semimetal requires a thin-film or cleaved-crystal geometry
\cite{xie2015new,schoop2016nodal}. The model assumes a field uniform along
$z$ and translationally invariant in $y$, whereas a surface stripe's fringe field 
decays into the bulk over a scale set by the stripe width. 
The idealization therefore holds when the film thickness is smaller 
than this decay length, so that all carriers experience a common $B_z$. 
Stripe and barrier widths $L \sim 25-400$\,nm  (i.e. $L \sim 1-15$ in our dimensionless units at $1$\,T)
are compatible with current nanofabrication techniques, although a quantitative analysis 
of the fringe-field profiles would be required to model a realistic experimental setup.
As we show in Sec.~\ref{Sec:Transport}, the two channels
are extinguished at $BL = 2q_0$ and $2\sqrt{q_0^2+E}$, respectively. 
For $q_0=53$, the lower threshold is $BL \approx 106$. 
For the widest stripes considered here, $L \approx 15$ ($\approx 390$~nm), 
this corresponds to $B \approx 7$~T, whereas narrower stripes require proportionally 
larger fields, for example $B\approx 21$~T for $L=5$. These estimates suggest 
that the required field–width product may be experimentally accessible. 
Whether a given stripe geometry and magnetization can generate the assumed 
localized field profile, however, requires a realistic micromagnetic calculation, 
which lies beyond the scope of the present work.

\begin{figure}[t]
    \centering
    \includegraphics[width=0.5\textwidth]{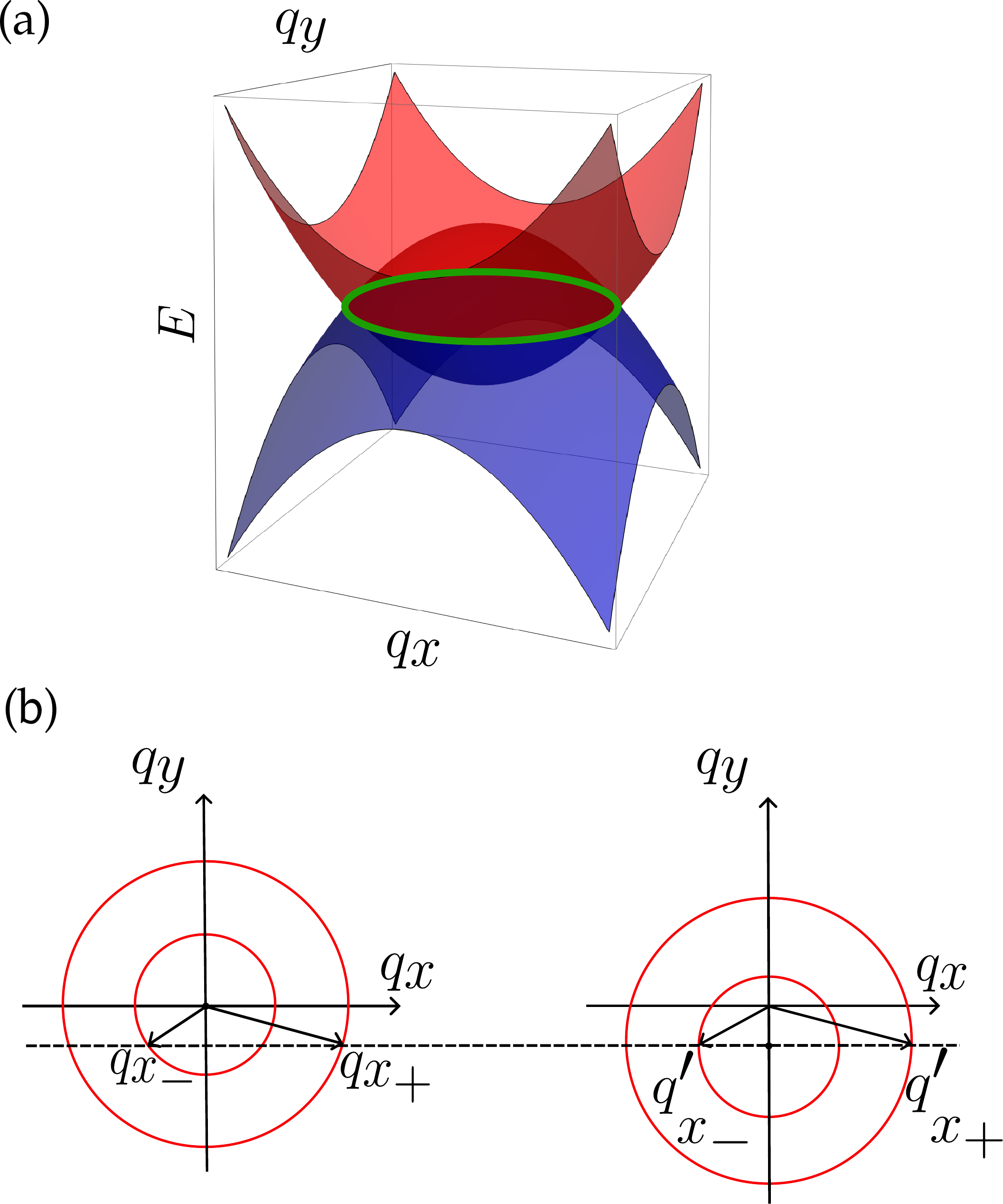}
    \caption{(a) Bulk dispersion \eqref{eq:dispersion} for $q_z=0$.
    (b)  Cross-sections of a constant-energy surface at fixed $q_z$ before the barrier (left panel) 
    and after the barrier (right panel). For fixed $(E,q_z)$, there are two propagating channels, 
    corresponding to the inner and the outer contour.
    The dashed line shows a fixed value of the conserved momentum $q_y$, and  
    $q_{x,\pm}$ are the magnitude of the allowed longitudinal component of the momentum 
    in the region $x<0$, see Eq.~\eqref{eq:qxs}.
    In the transmitted region $\tx>\tL$, the vector potential shifts the contour center 
    by $-\tB\tL$ along $q_y$, producing transmitted channels with different longitudinal momenta
    $q'_{x,\pm}$, see Eq.~\eqref{eq:qxsrt}.}
    \label{fig:FS}
\end{figure}

\section{Wave functions}
\label{Sec:Eigensystem}

In the presence of the magnetic barrier, translational invariance
in the $x$-direction is broken. We then label the states by
quantum numbers  $E$, $q_y$ and $q_z$.
In the field-free regions $x<0$ and $x>L$, for each $q_z$ the dispersion splits into two channels, 
corresponding to the inner and outer contours obtained by slicing the Fermi surface at fixed $q_z$, 
each channel carrying a well-defined group velocity. 
Inside the magnetic barrier ($0<x<L$), the Landau-gauge substitution maps the
Hamiltonian onto a shifted harmonic oscillator, whose solutions are
expressed in terms of parabolic-cylinder functions.

\subsection{Non-magnetic region}
\label{Sec:non-magnetic_region}

For $B=0$ the wave functions can be chosen as plane waves, $\psi(x) \propto e^{iq_x x}$, 
and the associated dispersion relation, illustrated in Fig.~\ref{fig:FS}(a), reads
\begin{equation}
\label{eq:dispersion}
    E= \pm \sqrt{(q_x^2+q_y^2-q_0^2)^2+\co^2 q_z^2},
\end{equation}
where the sign $\pm$ refers to conduction/valence bands.
The two bands cross at zero energy along the nodal ring $q_z=0$, $\sqrt{q_x^2+q_y^2}=q_0$. 
The scale $2q_0^2$ sets the maximal band separation in the inverted-band region and
is typically a fraction of eV~\cite{qiu2019observation,takane2018observation},
providing a natural ultraviolet cutoff for the effective model.

For $|E|<q_0^2$, the constant-energy surface has the topology of a torus. 
Any fixed-$q_z$ cross-section consists of two branches, 
described by
\begin{equation}
    q_x^2+q_y^2 = R_s^2(q_z),\qquad s=\pm,
    \label{eq:s_branch}
\end{equation}
which correspond to an outer ($s=+$) and inner ($s=-$) contour 
of radius 
\begin{equation} \label{eq:contourradius}
    R_s(q_z)=\sqrt{q_0^2+s \sqrt{E^2-c^2q_z^2}},
\end{equation}
with $R_+>R_-$, see Fig.~\ref{fig:FS}(b). We often refer to these branches as channels.
The two contours coalesce for $\co q_z=\pm E$.
On each contour, the longitudinal momentum magnitude is
\begin{equation}
 \label{eq:qxs}
    q_{x,s}=\sqrt{q_0^2-q_y^2+s\sqrt{E^2-\co^2 q_z^2}}.
\end{equation}
With this parameterization, 
the eigenstates of energy $E$ can be written as
\begin{equation}
    \psi_{s,r}(x)= \chi_s \, e^{irsq_{x,s}x}, \quad s=\pm 1, \quad r=\pm 1,
    \label{eq:psisr}
\end{equation}
where the normalized two-component spinors $\chi_s$ read
%
%
\begin{equation}
\label{eq:chi_s}
    \chi_s = \mathcal{N}_s
    \begin{pmatrix}
        i \co q_z\\
        s\sqrt{E^2-\co^2 q_z^2}-E
    \end{pmatrix} 
\end{equation}
with normalization 
\begin{equation}
\mathcal{N}_s =    \frac{1}{\sqrt{ 2E \left(E-s\sqrt{E^2-\co^2 q_z^2}\right)}}.
\end{equation}
%
The index $r=\pm$ in Eq.~\eqref{eq:psisr} relates to the propagation direction. 
Indeed, the group velocity of the state $\psi_{s,r}$ along $x$ takes the form $v_{s,r}=rv_s$, 
where
\begin{equation}
    v_{s}=\frac{2 q_{x,s} \sqrt{E^2-\co^2 q_z^2}}{E} >0 .  
    \label{eq:v_sr}
\end{equation}
The factor $rs$ in the exponent of Eq.~\eqref{eq:psisr} 
is chosen so that $r=+1$ labels right-moving states in both channels,
 independently of the channel index. Thus, states incident 
on the barrier from the left have $v_{s,r}>0$, or equivalently $r=+1$, 
whereas reflected states are left-moving and have $r=-1$.
	
\subsection{Magnetic region}
\label{Sec:magnetic_region}

In the barrier region $0<x<L$, the Hamiltonian~\eqref{eq:H_dimless} 
can be cast in the form
\begin{equation}
    \mathcal{H} = \co q_z\tau_y + 2 B \left( a^\dagger a - \Delta \right) \tau_z, 
    \label{eq:HB_ladder}
\end{equation}
where $\Delta=\frac{q_0^2}{2B}-\frac{1}{2}$, and we have defined the ladder operators
\begin{align*} 
    a & =\frac{1}{\sqrt{2B}}\!\left(\partial_x+q_y+Bx\right), \\
    a^\dagger & =\frac{1}{\sqrt{2B}}\!\left(-\partial_x+q_y+Bx\right), 
\end{align*}
with $\left[ a, a^\dagger \right]=1$. The operator $a^\dagger a$ is diagonalized by the
parabolic-cylinder functions $D_p(\pm\xi)$~\cite{gradshteyn2014table}, which
satisfy the identity
\begin{align}
    a^\dagger a \, D_p(\pm\xi) &= p \,  D_p(\pm\xi) ,
\end{align}
where $\xi=\sqrt{\frac{2}{B}}\,(Bx+q_y)$ is the rescaled and shifted coordinate.
Using the Ansatz $\psi(x) = \chi  D_p(\pm \xi)$ where $\chi$ is a two-component spinor, 
the eigenvalue equation for the Hamiltonian~\eqref{eq:HB_ladder} reduces to the algebraic problem
\begin{equation}
    \left[ \co q_z\tau_y + 2 B \left( p - \Delta \right) \tau_z -E \right]\chi=0.
    \label{eq:HB_ladderalge}
\end{equation}
The existence of a nontrivial solution requires a vanishing determinant:
\begin{equation}
 c^2q_z^2 +   4B^2 (p-\Delta)^2 -E^2=0.
\end{equation}
Solving for $p$, we find two branches
\begin{equation}
    p_s = \Delta + \frac{s}{2B} \sqrt{E^2-\co^2 q_z^2}, \quad s=\pm .
    \label{eq:ps}
\end{equation}
In an infinite system with spatially uniform field, wave function normalizability would require
$p_s$ to be a non-negative integer, giving $q_z$-dependent Landau levels.
In the present finite-width barrier, however,  $p_s$ can be generically non-integer,
and both  $D_{p_s}(\xi)$ and $D_{p_s}(-\xi)$ are allowed local solutions. 
Replacing Eq.~\eqref{eq:ps}  in Eq.~\eqref{eq:HB_ladderalge} and solving for $\chi$, 
one finds the same solutions as in Eq.~\eqref{eq:chi_s}.
The eigenstates of Eq.~\eqref{eq:HB_ladder} can then be cast in the form
\begin{equation}
    \psi_{s,r}(x) = \chi_s  D_{p_s}(r\xi), \quad s=\pm, \quad r=\pm,
    \label{eq:pc_wave}
\end{equation}
with $\chi_s$ given in Eq.~\eqref{eq:chi_s}.
We note that the condition \eqref{eq:ps} coincides with Eq.~\eqref{eq:s_branch} with 
$q_x^2+q_y^2-q_0^2$ replaced by $2B(p_s-\Delta)$, and the branch index $s$ in Eq.~\eqref{eq:ps}
can be identified with the branch index in Eq.~\eqref{eq:s_branch}.

Because the branch index $s$ labels the same spinor $\chi_s$ throughout the system,  
the two branches can be tracked consistently across the barrier; 
we show in Sec.~\ref{Sec:scattering_amplitudes} that they decouple, 
so each channel can be solved separately.

\section{Scattering and conductance}
\label{Sec:Transport}

Having established the local solutions in the three regions, we now turn to
the scattering problem. We begin by deriving a general kinematic criterion 
for nonvanishing transmission in terms of the barrier strength $\tB\tL$. 
We then construct the full scattering solution by matching wave functions 
and their derivatives at the two barrier edges, obtaining closed-form expressions 
for the channel-resolved transmission and reflection amplitudes in terms of parabolic-cylinder
functions. The conductance follows by integrating the total transmission probability
over transverse momenta. We close the section by discussing the 
numerical results for the transmission probabilities, their angular dependence,
and the conductance as functions of magnetic field, barrier width, and
energy.

\subsection{Threshold condition for transmission}
\label{Sec:Critical}

Since outside the magnetic barrier the constant-energy surface has toroidal geometry,
it is convenient to parameterize the momentum of incident states at fixed energy $E>0$ 
in terms of two angles as
%
\begin{align}
    q_{x} &=  \sqrt{q_0^2 + E\sin\theta}\,\cos\phi, \nonumber\\
    q_{y} &= \sqrt{q_0^2 + E\sin\theta}\,\sin\phi, \label{eq:parametric1}\\
    q_{z} &= \frac{E}{\co}\cos\theta. \nonumber
\end{align}
with $\theta \in \left[0,2\pi\right)$ and $\phi\in \left(-\pi/2,\pi/2\right)$.
Comparing Eq.~\eqref{eq:parametric1} with Eq.~\eqref{eq:s_branch}, we see that outer contours ($+$ branch) 
correspond to $\theta$ in the range $\left(0,\pi \right)$, while inner contours ($-$ branch)
to $\theta$ in the range $\left(\pi,2\pi \right)$. The magnetic barrier deflects the electron trajectory 
in the $xy$-plane and in the transmitted region $x>L$,
we have  
\begin{align}
    q'_{x} &= s \sqrt{q_0^2 +  E\sin\theta}\,\cos\phi', \nonumber\\
    q_{y} +  BL &= \sqrt{q_0^2 +  E\sin\theta}\,\sin\phi', \label{eq:parametric2}\\
    q_{z} &= \frac{E}{\co}\cos\theta,  \nonumber
\end{align}
where translational invariance in the $y$ and $z$ directions 
implies conservation of the canonical momenta $q_y$ and $q_z$. 
Conservation of $q_z$ does not, by itself, exclude the transformation $\theta \to 2\pi-\theta$ 
in the transmitted region. Such a transformation would correspond to transmission into the opposite 
contour after propagation through the magnetic barrier, and is kinematically allowed. 
As shown by the explicit matching calculation below, however, the inter-contour transmission amplitude vanishes.

From Eqs.~\eqref{eq:parametric1} and \eqref{eq:parametric2}, we obtain
\begin{equation}
    \sin\phi' = \sin\phi + \frac{\tB\tL}{\sqrt{q_0^2 + E\sin\theta}},
    \label{eq:phi_shift}
\end{equation}
which describes a transverse momentum kick imparted by the magnetic
barrier. Since $|\sin\phi'|\leq 1$, Eq.~\eqref{eq:phi_shift} implies that an
incident state is completely reflected once the barrier strength $BL$
exceeds the angle-dependent critical value
\begin{equation}
    \tB\tL =\left( 1-\sin\phi \right)\sqrt{q_0^2 + E\sin\theta}\,,
    \label{eq:critical_BL}
\end{equation}
at which the transmitted longitudinal momentum turns imaginary. This is the threshold
for an individual scattering state. In particular, at normal incidence ($q_y=q_z=0$), 
the channel-$s$ critical value becomes $\sqrt{q_0^2+sE}$.

Maximizing the threshold in Eq.~\eqref{eq:critical_BL} over the angles
sets the field beyond which a channel is extinguished entirely. The
factor $(1-\sin\phi)$ is maximal at grazing incidence $\sin\phi=-1$
for both channels, so the channel distinction lies in the radial
factor $\sqrt{q_0^2+E\sin\theta}$. For the $+$ channel
($\theta\in(0,\pi)$, $\sin\theta > 0$) this is largest at
$\theta=\pi/2$, giving $BL = 2\sqrt{q_0^2+E}$. For the $-$ channel
($\theta\in(\pi,2\pi)$, $\sin\theta\le 0$) it is instead largest at
the edges of the range, $\sin\theta\to 0^-$, giving $BL = 2q_0$. 
Since $2q_0 < 2\sqrt{q_0^2+E}$, total reflection sets in first in the $-$
channel and only later in the $+$ channel as $BL$ increases.
This sequential quenching governs the two-terminal conductance, 
which we analyze in Sec.~\ref{Sec:Conductance}.

\subsection{Channel-resolved scattering probabilities}
\label{Sec:scattering_amplitudes}

We now solve in detail the one-dimensional scattering problem at energy $E>0$ 
and fixed $q_y$ and $q_z$.
Before constructing the scattering solution, we show that the two
channels decouple exactly. In each of the three regions the state is a
superposition $\psi(x)=\sum_{s,r} a_{s,r}\,\psi_{s,r}(x)$ of the local
solutions of Sec.~\ref{Sec:Eigensystem}, and both solutions $r=\pm$ 
of a given branch share the same spinor $\chi_s$
[Eq.~\eqref{eq:chi_s}]. The state is therefore expressed as
$\psi(x)=\chi_+ f_+(x)+\chi_- f_-(x)$, with $f_s(x)$ collecting the
$r=\pm$ factors of branch $s$. The spinors $\chi_+$ and $\chi_-$ are
independent of $x$ and identical in all three regions; for
$cq_z\neq\pm E$ they are linearly independent and form a fixed basis of
the spinor space. Continuity of $\psi$ and $\partial_x\psi$ at $x=0$ and
$x=L$ then equates two spinors, each already expanded in this basis, and
since the expansion of a vector in a basis is unique the coefficients of
$\chi_+$ and of $\chi_-$ must match separately. This equality of
coefficients resolves the matching into two independent
scalar problems, one per branch. No amplitude connects $s=+$ to $s=-$:
the inter-contour transmission and reflection amplitudes vanish
identically, and the branch index $s$ is conserved throughout. We may
therefore solve each channel separately. 
The single exception is the measure-zero set $cq_z=\pm E$, where the two contours
coalesce ($q_{x,+}=q_{x,-}$), $\chi_+=\chi_-$, and the basis degenerates.

In the incident region ($\tx<0$), magnetic barrier ($0<\tx<\tL$), and transmitted region ($\tx>\tL$), 
the wave functions for channel $s$ are
\begin{align}
    \psi_{\mathrm{in}}^{s}(\tx) &=
        \frac{\chi_s}{\sqrt{v_s}} \left[  
        e^{isq_{x,s}\tx} + r_s  \, e^{-isq_{x,s}\tx} \right],\\
    \psi_{\tB}^{s}(\tx) & =\chi_s  
    \left[ a_{s,1}  D_{p_s}(\xi) + a_{s,2} D_{p_s}(-\xi) \right],\\
    \psi_{\mathrm{out}}^{s}(\tx) &=
         \frac{\chi_s}{\sqrt{v'_s}} t_s \,
        e^{isq'_{x,s}(\tx-\tL)},
\end{align}
where $a_{s,i}$ are complex amplitudes, the incident momentum $q_{x,s}$ 
is given in Eq.~\eqref{eq:qxs} and the transmitted one is
\begin{equation}
\label{eq:qxsrt}
    q'_{x,s} =\sqrt{q_0^2-(q_y+\tB\tL)^2+s\sqrt{E^2-\co^2q_z^2}}.
\end{equation}
The velocity factors are included so that the states are normalized to unit flux, and the 
coefficients $r_s$ and $t_s$ are the reflection and transmission amplitudes. 
The velocity of the transmitted state, $v_s'$, is obtained 
from Eq.~\eqref{eq:v_sr} by replacing $q_{x,s} \to q_{x,s}'$.

Since the Hamiltonian is second order in $\partial_{\tx}$, both $\psi$
and $\partial_{\tx}\psi$ must be continuous at $\tx=0$ and $\tx=\tL$.
Matching at the two interfaces yields, for each channel, a $4\times4$
linear system for the coefficients $r_s, a_{s,1}, a_{s,2}, t_s$~\cite{ibrahim1995two}.
To write the solution compactly, we define the boundary values of the
coordinate $\xi$:
\begin{equation}
    \xi_0=\sqrt{\frac{2}{\tB}}\, q_y,\qquad
    \xi_L=\sqrt{\frac{2}{\tB}}\,(\tB\tL+q_y),
\end{equation}
and the auxiliary combinations ($\beta=\pm$)
\begin{align}
    u_{1,s}^{\beta} & =  D'_{p_s}(-\xi_L)
        + \beta i\, D_{p_s}(-\xi_L)\, q'_{x,s}, \\
    u_{2,s}^{\beta} & =  D'_{p_s}(\xi_L)
        + \beta i\, D_{p_s}(\xi_L)\, q'_{x,s}, \\
    v_{1,s}^{\beta} & =  D'_{p_s}(-\xi_0)
        + \beta i\, D_{p_s}(-\xi_0)\, q_{x,s}, \\
    v_{2,s}^{\beta} & =  D'_{p_s}(\xi_0)
        + \beta i\, D_{p_s}(\xi_0)\, q_{x,s},
\end{align}
where we use a prime to indicate the derivative with respect to $x$ 
(not with respect to the argument): $D_p'(\xi) = \partial_x D_p(\xi)$.
The transmission and reflection amplitudes are then
\begin{equation}
    t_{s} = 2 \sqrt{\frac{v'_s}{v_s}} \frac{q_{x,s} 
    \left[D'_{p_{s}}(\xi_L)  D_{p_{s}}(-\xi_L) - D'_{p_{s}}(-\xi_L) D_{p_{s}}(\xi_L)\right] }
        {i \left( v_{2,s}^{+}\, u_{1,s}^{-}
        - v_{1,s}^{+}\, u_{2,s}^{-} \right)}, \label{eq:t_s} 
\end{equation}
%
and
\begin{equation}
    r_{s} =s \frac{u_{2,s}^{\bar s}\, v_{1,s}^{\bar s} - u_{1,s}^{\bar s}\, v_{2,s}^{\bar s}}
        {u_{1,s}^{\bar s}\, v_{2,s}^{s} - u_{2,s}^{\bar s}\, v_{1,s}^{s}},  
        \label{eq:r_plus} 
\end{equation}
with $\bar s = -s$.
%
The transmission and reflection probabilities for the two channels follow 
as $|t_s|^2$ and $|r_s|^2$ and we checked that they satisfy the unitarity 
condition separately in each channel:
\begin{equation}
    |t_s|^2 + |r_s|^2=1.
\end{equation}

\begin{figure}[ht]
	\centering
	\includegraphics[width=0.48\textwidth]{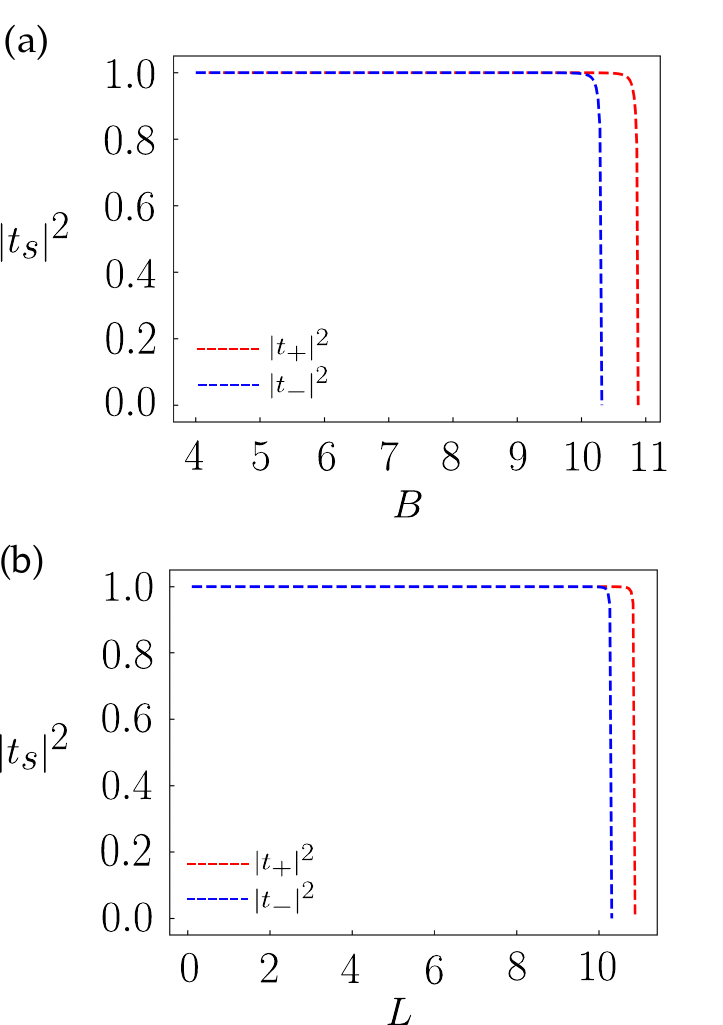}
	\caption{Transmission probabilities $|t_s|^2$ at normal incidence ($q_y=q_z=0$)
     for energy $E=150$ ($10.5$~meV), plotted as functions of magnetic field $B$ with $\tL=5$ ($130$~nm), 
     (panel (a)) and of barrier width $L$ with $B=5$ (panel (b)). 
  The transmission in channel $s$  abruptly vanishes when $\tB\tL$ reaches the critical value $\sqrt{q_0^2+sE}$ (for the inner and outer contour, nearly $51.6$ and $54.4$, respectively) beyond which $q'_{x,s}$ becomes imaginary. Here the nodal-ring radius is $q_0=53$ and $c=148$.}
	\label{fig:trans_vs_B}
\end{figure}

Figure~\ref{fig:trans_vs_B}(a) displays the transmission probabilities
$|t_s|^2$ at normal incidence ($q_y = q_z = 0$) and energy $E=150$,
as functions of the applied magnetic field $\tB$ at fixed barrier width
$\tL=5$ and nodal-ring radius $q_0=53$. At weak field the transmission 
remains close to unity over most of the propagating range 
and then falls sharply to zero near the kinematic threshold. 
The key observation is that the two channels are suppressed at different
critical fields: the $-$ channel (blue curve) drops to zero first, at
$\tB\tL = \sqrt{q_0^2 - E}$, while the $+$ channel (red curve) survives
to higher field and vanishes at $\tB\tL = \sqrt{q_0^2 + E}$. Both values
follow from Eq.~\eqref{eq:critical_BL} at $\phi=0$, with $\theta=3\pi/2$
and $\theta=\pi/2$ for the $-$ and $+$ channels respectively. 
These thresholds
are specific to normal incidence; electrons incident at an angle can
absorb a larger transverse shift, so each channel as a whole keeps
conducting to higher field, as we discuss in
Sec.~\ref{Sec:Conductance}.

Figure~\ref{fig:trans_vs_B}(b) shows the complementary case of 
varying the barrier width $\tL$ at fixed field $\tB = 5$, with the 
same energy and nodal-ring parameters.
The behavior mirrors panel~(a): both channels begin with near-unit 
transmission at small $\tL$ and fall sharply to zero as $BL$
approaches the kinematic threshold, with the $-$ channel transmission again vanishing before the $+$ channel.
This is expected since for $q_y = q_z = 0$, the critical 
barrier widths are $\tL = \sqrt{q_0^2 \pm E}/\tB$ for the two channels, respectively.
 
The asymmetric suppression of the two channels can be understood from 
the structure of the available phase space for transmission.
At a given energy, the inner contour encloses a smaller 
area in the $q_xq_y$-plane than the outer contour, 
and correspondingly the outgoing momentum $q'_{x,-}$ is smaller in 
magnitude than $q'_{x,+}$ [see Fig.~\ref{fig:FS}(b)].
A smaller $q'_{x,s}$ means the channel is closer to its cutoff: the 
point at which the transverse shift $\tB\tL$ in $q_y$ pushes the 
momentum outside the Fermi contour and $q'_{x,s}$ turns imaginary.
By tuning $\tB\tL$ into the window $2q_0< \tB\tL< 2\sqrt{q_0^2 + E}$
one can suppress the inner-contour channel entirely while leaving 
the outer-contour channel partially transmitting. In our case for Figure~\ref{fig:G_vs_B}(a), $\Delta B \approx 3.5 $ T, which is the window where the inner-contour is completely blocked. This channel-resolved filtering has no analogue in Weyl or 
Dirac semimetals, where only a single propagating mode exists at 
each node.

 \begin{figure}[t]
	\centering
	\includegraphics[
    width=.4\textwidth]{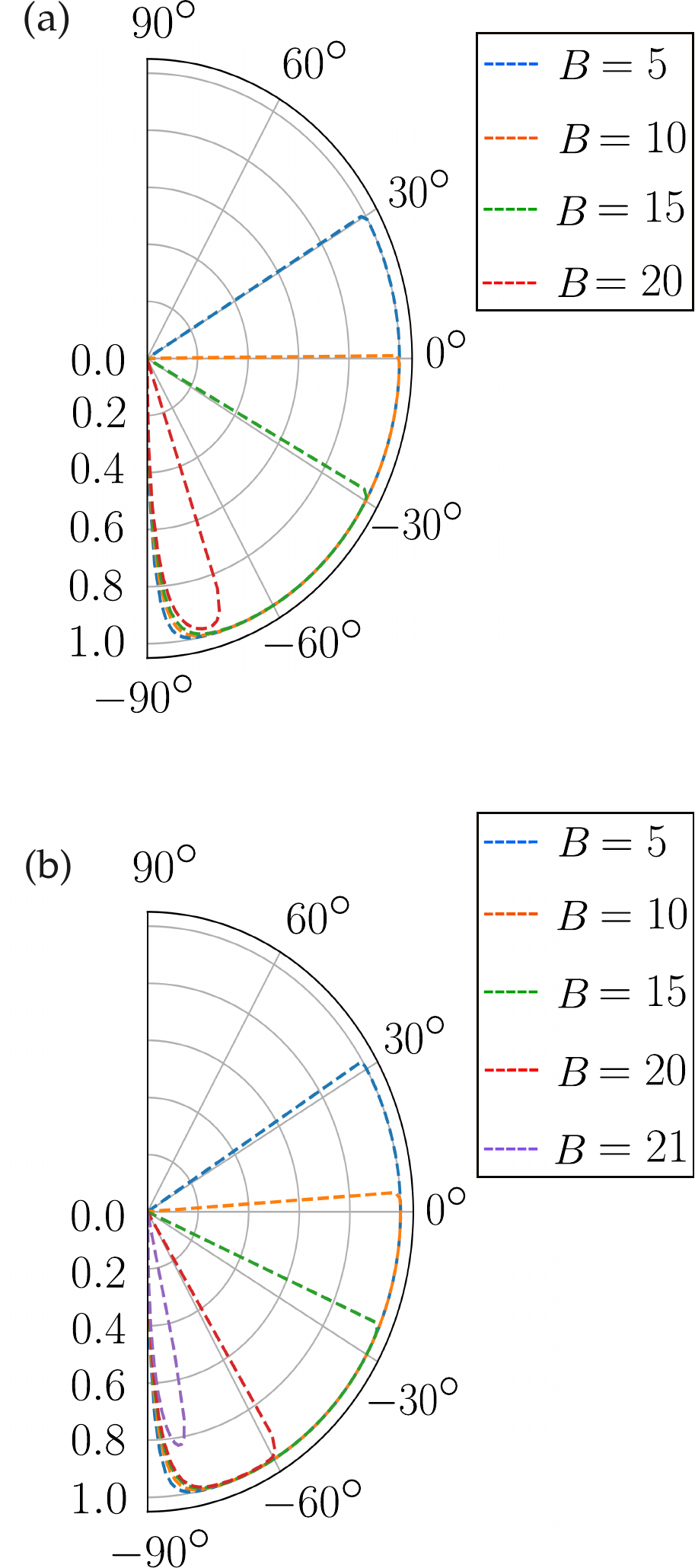}
    \caption{
    Polar plots of transmission probabilities at $q_z=0$ as a function of the azimuthal angle $\phi$, 
    for channel $-$ [panel (a)] and channel $+$ [panel (b)], for several values of the magnetic field. 
    We use the same parameters as in Fig.~\ref{fig:trans_vs_B}(a):  
    $E=150~(10.5$~meV), $\tL=5$ ($130$~nm), $q_0=53$, $c=148$.
    }
	\label{fig:polar}
\end{figure}

\subsection{Angular dependence}
\label{Sec:angdep}

Figure~\ref{fig:polar} shows polar plots of the transmission
probability at energy $E=150$ as a function of the azimuthal angle $\phi$ at $q_z=0$,
for the $-$ channel [panel~(a)] and the $+$ channel
[panel~(b)], for several values of the magnetic field.
On this $q_z$ slice the relevant contour radius is
$R_s(0)=\sqrt{q_0^2+sE}$ with $R_+(0)>R_-(0)$.
 A propagating transmitted state exists only where $|\sin\phi'|\le 1$; since $BL>0$,
the binding constraint is the upper one,
$\sin\phi\le 1-BL/R_s$, while the lower bound $\sin\phi\ge-1$ is
unaffected. The angular acceptance is therefore the asymmetric window
$\phi\in\left[-\tfrac{\pi}{2},\,\arcsin\left(1-BL/R_s\right)\right]$. 
In the limit
$BL\to 0$ it approaches the full range $(-\pi/2,\pi/2)$; already at
the weakest field shown ($B=5$, $BL=25$) the upper edge has retreated
to $\phi$ around  $30^\circ$, and the lobe is visibly one-sided.
As $B$ increases, the upper edge $\arcsin\!\left(1-BL/R_s\right)$ sweeps
downward through zero and into negative angles, so the lobe contracts
onto the $\phi<0$ side. 
Transmission persists longest for grazing incidence $\phi \to -\pi/2$,
where the initial transverse momentum lies near the lower edge of the
Fermi contour: there $1 - \sin\phi$ is maximal, so the barrier's
positive shift in $q_y$ is most easily accommodated [cf.~Eq.~\eqref{eq:critical_BL}], 
leaving the largest available range for the shift.
The $q_z=0$ window closes entirely at $BL=2R_s(0)=2\sqrt{q_0^2+sE}$, i.e.\ at
$B\approx 20.6$ for the $-$ channel and $B\approx 21.8$ for the $+$
channel, consistent with panels~(a) and~(b).

Because $R_+>R_-$, the outer channel collimates more slowly and survives 
to higher field, so the two contours can be distinguished 
by the field at which their angular acceptance collapses.
The progressive, asymmetric narrowing in Fig.~\ref{fig:polar} shows that
angle-resolved transport could  probe the two-channel
Fermi-surface geometry, since both the collimation rate and the
direction of maximal transmission ($\phi=-\pi/2$) are set by the contour
geometry, and the closing fields differ between channels.

\subsection{Conductance}
\label{Sec:Conductance}

The two-terminal conductance per unit transverse area is obtained by integrating the total
transmission over the  transverse momenta~\cite{datta1997,matulis1994wave}  and reads
\begin{equation}
    \frac{G}{A} =\frac{e^2}{h}\frac{1}{4\pi^2\ell_0^2} 
    \iint dq_y dq_z \sum_{s=\pm} |t_{s}|^2 .
    \label{eq:conductance}
\end{equation}
Equation \eqref{eq:conductance} gives the conductance per spin. 
If spin degeneracy is retained, the result should be multiplied by $g_{\rm spin}=2$. 
The integration is restricted to the region of the  $q_yq_z$ plane in which
$q_{x,s}$ and $q'_{x,s}$ are both real. On the incident side this requires
\begin{equation}
    |q_y| \le R_s(q_z),
    \label{eq:qy_in}
\end{equation}
while on the transmitted side
\begin{equation}
    |q_y+BL| \le R_s(q_z),
    \label{eq:qy_out}
\end{equation}
with $R_s(q_z)$ the $q_z$-fixed contour radius defined in Eq.~\eqref{eq:contourradius}
and $q_z\in[-E/\co,\, E/\co]$. The conducting phase space for channel
$s$ is the overlap of these two intervals, which is given by
\begin{equation}
\label{eq:overlap_qy}
    \Delta q_{y,s} (q_z)= {\rm max} \left[ 0, 2R_s(q_z)-BL\right].
\end{equation}
Therefore, with $|t_s|^2 \simeq 1$, we arrive at 
\begin{equation}
    \frac{G}{A} \simeq \frac{e^2}{h}\frac{1}{4\pi^2\ell_0^2} 
    \int^{E/c}_{-E/c} dq_z \sum_{s=\pm} {\rm max} \left[ 0, 2R_s(q_z)-BL\right].
    \label{eq:conductance_est}
\end{equation}
As $BL$ increases, the overlap shrinks linearly, and the channels close one by one: complete 
suppression occurs at $BL = 2q_0$ for the $-$ channel and $BL = 2\sqrt{q_0^2+E}$ for the $+$ channel,  
in agreement with the discussion in Sec.~\ref{Sec:Critical}.

\begin{figure}[h]
		\centering
		\includegraphics[width=0.5\textwidth]{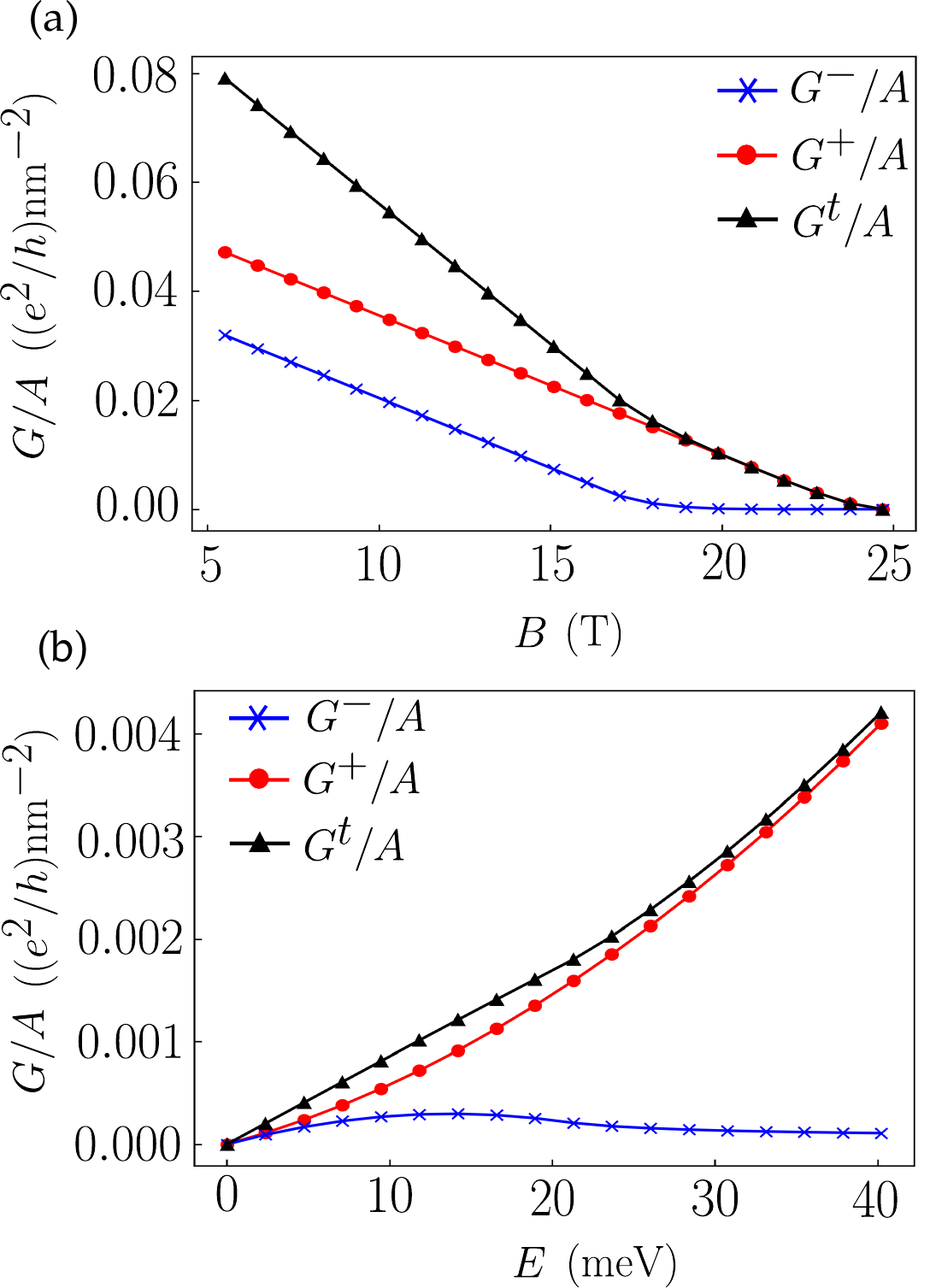}
		\caption{      
        (a) 
        Two-terminal conductance per unit area $G/A$, Eq.~\eqref{eq:conductance}, 
        in units of $(e^2/h)/nm^2$ at energy $E=1000$ ($70$~meV) as a function of  
        magnetic field $\tB$ for $L=5$ ($130$~nm), $q_0=53$, $c=148$. 
        The conducting phase-space is shrinking linearly with increasing $B$ [see Eqs.~\eqref{eq:qy_in}--\eqref{eq:qy_out}] 
        and the $-$ channel stops conducting before the $+$ channel.  
        (b) Conductance vs energy $E$ at  $\tL=20$ ($\sim 520$~nm) for $\tB=5$ T and $q_0=53$. 
        At higher energy, most of the conductance is carried by the $+$ channel. 
        Although the transmission probability $|t_-|^2\sim 1$ for the $-$ channel, its conductance \emph{decreases} 
        with increasing $E$ because the allowed $q_y$ range shrinks faster than the $q_z$ range expands [see Eqs.~\eqref{eq:qy_in}--\eqref{eq:qy_out}]. 
        For the $+$ channel, both ranges expand with $E$, leading to increasing conductance.} 
		\label{fig:G_vs_B}
	\end{figure}

Figure~\ref{fig:G_vs_B}(a) shows the conductance versus magnetic
field $B$. Three curves are displayed: $G^+$ (outer channel, red),
$G^-$ (inner channel, blue), and the total $G^t = G^+ + G^-$ (black).
Both channels decrease approximately linearly with $B$, following the
linear shrinkage of the conducting width $\Delta q_{y,s}$
[Eq.~(\ref{eq:overlap_qy})]. Because $2q_0 < 2\sqrt{q_0^2+E}$, the $-$
channel is extinguished first: the $G^-$ curve reaches zero before
$G^+$, and in the window between the two thresholds only the outer
Fermi contour contributes to magneto-transport.

Figure~\ref{fig:G_vs_B}(b) shows the conductance versus energy
$E$, where the two channels again behave differently. For the outer
channel (red curve), $G^+$ rises monotonically: increasing $E$ widens both
the longitudinal window $|q_z| \le E/c$ and the contour radius
$R_+(q_z) = \sqrt{q_0^2 + \sqrt{E^2 - c^2 q_z^2}}$, so the available
phase space grows along both axes. The inner channel (blue curve) 
instead rises at small $E$, turns over, and falls. Here the two extents work
against each other: the $q_z$ window still widens with $E$, but the
inner contour radius $R_-(q_z) = \sqrt{q_0^2 - \sqrt{E^2 - c^2 q_z^2}}$
shrinks. Geometrically, raising $E$ inflates the toroidal Fermi
surface: the outer contour expands while the inner one contracts
toward the ring axis, closing the central hole of the torus, with
$R_-(0) = \sqrt{q_0^2 - E}$ vanishing at the edge of the toroidal
regime $E = q_0^2$. At small $E$ the widening $q_z$ window dominates
and $G^-$ grows; beyond a crossover the contracting transverse phase
space wins and $G^-$ falls, even though the per-mode transmission
$|t_-|^2$ stays close to unity. The total conductance $G^t$ (black curve) 
is therefore carried almost entirely by the outer channel at high energy.

We emphasize that a two-terminal measurement returns only the total $G^t$; 
the channel-resolved curves $G^\pm$ are a theoretical decomposition 
guaranteed by conservation of $s$ in the scattering process but not 
separately measurable. In $G^t$ the two channels' contributions merge 
into a single smooth decrease, terminating once the outer channel is cut off.

\section{Conclusions}
\label{Sec:Conclusions}

We have studied quantum transport through a magnetic barrier in a
nodal-line semimetal, in the regime where the Fermi energy lies close
to the nodal ring and each fixed-$q_z$ cross-section of the toroidal
Fermi surface supports two transport channels. Using a two-band
effective Hamiltonian and matching parabolic-cylinder wave functions
across the barrier, we obtained closed-form, channel-resolved
transmission and reflection amplitudes.

The central result is that the barrier resolves the inner and outer
contours: enclosing different momentum-space areas, they accommodate
the field-induced transverse-momentum kick unequally, so the inner
channel is cut off at a weaker field than the outer. In the window
$2q_0 < BL < 2\sqrt{q_0^2+E}$ the inner contour is fully blocked while
the outer still transmits, and the barrier acts as a channel-selective
filter.  This shapes the two-terminal conductance, which decreases as 
the inner channel closes near $BL=2q_0$ and vanishes once the outer 
channel is cut off at $BL=2\sqrt{q_0^2+E}$.
The energy dependence further distinguishes the contours: the outer-channel
conductance grows monotonically with $E$, whereas the inner-channel
conductance turns over and decreases as its transverse phase space
shrinks, despite a per-mode transmission that stays near unity.
Together with the channel-dependent angular collimation, these
features are experimentally accessible fingerprints of the toroidal
Fermi surface and the nodal ring that produces it.

Several directions deserve further study. Experimentally, the barrier
could be realized with a ferromagnetic stripe on a thin film or cleaved
crystal of a clean NLS such as Ca$_3$P$_2$, with the cutoff and
sequential quenching probed by two-terminal transport versus field and
gate-tuned Fermi energy; a quantitative comparison would require
modelling the realistic fringe field rather than the step barrier used
here. Theoretically, smooth edges, disorder, and finite temperature
would each soften the sharp cutoffs; the extent to which the sequential
quenching survives them remains to be quantified. Extensions to
oscillating fields and magnetic superlattices would further test how
robustly the multi-contour transport encodes the nodal-ring geometry.

\bibliographystyle{apsrev4-2}

\bibliography{references}

\end{document}